# Pocket Schlieren: a background oriented schlieren imaging platform on a smartphone

DIGANTA RABHA[1], VIMOD KUMAR[1], AKSHAY KUMAR[1], DINESH SAINI[2], AND MANISH KUMAR[1*]

## AFFILIATIONS

[1]*Centre for Sensors, Instrumentation and Cyber-physical System Engineering (SeNSE), Indian Institute of Technology Delhi, Hauz Khas, New Delhi-110016, India*

[2]*Optics and Photonics Centre (OPC), Indian Institute of Technology Delhi, Hauz Khas, New Delhi-110016, India*

*Corresponding author: kmanish@iitd.ac.in

**Abstract:**

Background-oriented schlieren (BOS) is a powerful technique for flow visualization. Nevertheless, the widespread dissemination of BOS is impeded by its dependence on scientific cameras, computing hardware, and dedicated analysis software. In this work, we aim to democratize BOS by providing a smartphone based scientific tool called "Pocket Schlieren". Pocket Schlieren enables users to directly capture, process, and visualize flow phenomena on their smartphones. The underlying algorithm incorporates *consecutive frame subtraction* (CFS) and *optical flow* (OF) techniques to compute the density gradients inside a flow. It performs on both engineered and natural background patterns. Using Pocket Schlieren, we successfully visualized the flow produced from a burning candle flame, butane lighter, hot soldering iron, room heater, water immersion heating rod, and a large outdoor butane flame. Pocket Schlieren promises to serve as a frugal yet potent instrument for scientific and educational purposes. We have made it publicly available at doi: 10.5281/zenodo.10949271.

**Introduction:**

Flow visualization in transparent media is an important method in aerodynamics, fluid mechanics, combustion, and related sciences [1]. A physical flow induces changes in optical density (or refractive index) resulting in bending of light rays. A measure of this bending forms the basis for optical flow visualization. Among various flow visualization techniques [2], schlieren imaging-based techniques are particularly interesting owing to their simplicity and sensitivity [3]. However, schlieren technique requires precision mirrors which get prohibitively expensive with increasing size thus greatly limiting the achievable field of view [4]. A variant of schlieren, called Background Oriented Schlieren (BOS), can visualize a very large-scale flow phenomenon [5]. BOS hardware setup consists of a patterned background, a camera, and a light source. The actual density gradient visualization happens through a digital image analysis software [6].

BOS has been extensively applied to visualize a variety of optical density gradient fields. In 2000, Dalziel *et al.* [5] and Raffel *et al.* [7] demonstrated the application of the BOS in flow visualization. Since then, various reports have been published showcasing the capability of BOS in flow field visualization and measurements, such as the density gradients induced by flames or plumes [8], shockwaves [9], supersonic jet [10], and vortices [11]. BOS has also been expanded to tomography setup for 3D imaging [12–14]. Using natural background (such as vegetation field, desert, trees etc.) has allowed to visualize helicopter rotor vortex and aircraft shockwave [15]. Similarly, BOS imaging using large natural celestial objects [16], and manmade retroreflective surfaces [17] have also been reported.

Significant efforts have been dedicated towards enhancing the accessibility of BOS hardware and software. Advanced applications involving the visualization of supersonic flow can be carried out using basic background patterns and lighting but necessitate the use of high-speed cameras. However, the camera hardware requirements can be significantly reduced for a general-purpose flow visualization application [18]. Nevertheless, the software requirements remain the same regardless of the camera hardware in use. Dedicated particle image velocimetry software e.g. PIVlab [19,20] is widely used for BOS reconstruction [8].

Smartphones, with a staggering 6.4 billion users worldwide [21,22], serve as an optimal tool for democratizing science. Modern smartphones, with powerful sensors and cameras, are highly capable scientific instruments for imaging and sensing [23–25]. They have also been used for BOS imaging applications [18,26]. However, these works used smartphones only for capturing images which were then processed on a desktop with dedicated high-end software like PIVlab to obtain flow visualization. A lot needs to be done to enable BOS software accessibility by eliminating its dependency on computers. Given the computational abilities of smartphones, it is desirable to extend their application to include BOS software leading to on-device flow visualization.

This paper presents a frugal scientific tool called “Pocket Schlieren” which is a smartphone based complete BOS system for flow visualization. Pocket Schlieren is essentially an Android app which can capture, process, and visualize BOS data. It incorporates two modes of BOS visualization 1) Live BOS, and 2) Pre-recorded BOS. It is freely downloadable from the following doi: 10.5281/zenodo.10949271. Below we provide more details and demonstrate a variety of flow visualization results using our tool. Pocket Schlieren may find applications in preliminary flow visualization laboratory experiments, gas leaks detection, cool flame detection, citizen science, and STEM education.

# 1. BOS Principle

## 1.1. Basic theory

BOS imaging working principle can be represented by simplified schematics as shown in Fig. 1a. Here, the density gradient field forms the subject under investigation. It is placed between a camera system on one side, and a patterned background on the other side. Here the camera system consists of an image sensor and a lens system. The background pattern is either self-luminous (on a display screen) or illuminated externally by a light source. The camera is focused to sharply image the background. When a light-ray from the background passes through the density gradient medium, it gets deflected with a deflection angle $\alpha$ due to refraction from the inhomogeneous refractive index in the medium. This deflection causes a local displacement ($\delta x_{bg}$) in the observed background pattern. This local displacement is a function of refractive index distribution in the media [27]. Thus, a measure of this displacement maps the refractive index variation leading to the flow visualization.

There are two methods to select image pairs for displacement calculations. The first method takes an image without the subject under investigation as a reference frame and compares it to the subsequent captures in the presence of the subject/flow (see Fig. 1b). In the second approach, there is no fixed reference frame. Instead, two consecutive frames form the image pair as shown in Fig. 1c. The displacement in background pattern plane, $\delta x_{bg}$ is directly related to the small deflection angle, $\alpha$ through $\delta x_{bg} = d_2\,\alpha$ [28]. Here, $d_2$ is the distance between background and the object under investigation. Knowing the imaging lens focal length $f$, and background to lens distance $d_3$, we can easily obtain the magnification between the background pattern plane and the sensor plane. Thus, displacement in the sensor plane becomes $\delta x_{sensor} = d_2\,\alpha\, f/(d_3 - f)$.

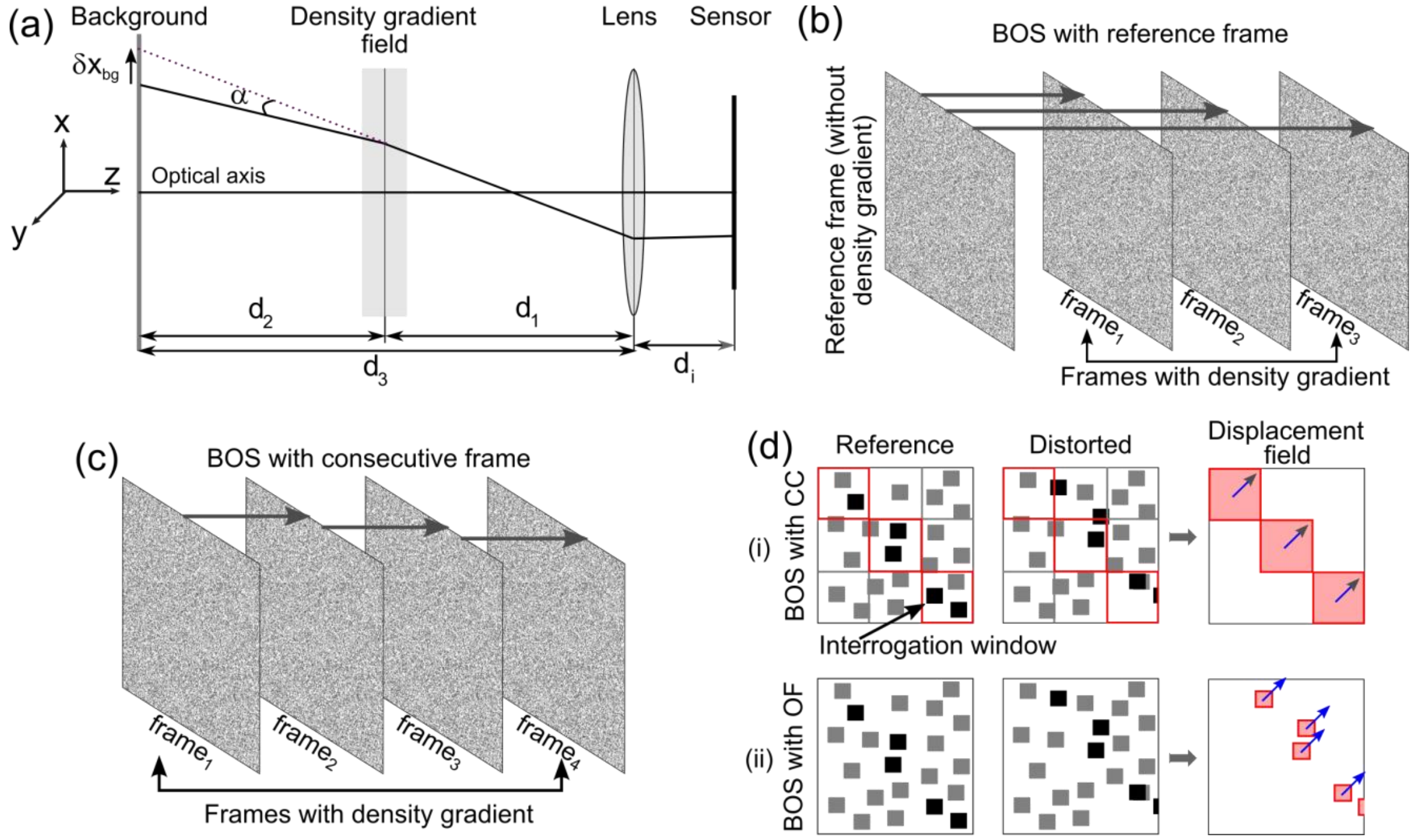

**Fig. 1:** BOS theory. (a) A simple schematic of the BOS principle, (b) BOS measurement with respect to a reference frame, (c) BOS measurement with consecutive frames, and (d) illustration of measurement of the displacement vectors for two popularly reported reconstruction techniques: cross correlation-BOS and optical flow-BOS.

The density gradient inside a flow gets captured by the deflection angle $\alpha = \frac{1}{n_0}\int \frac{\partial n}{\partial x} dz = \frac{Z}{n_0}\frac{\partial n}{\partial x}$ [4,5]. Here, $n_0$ = 1.000292 is the refractive index of the ambient air, n is the refractive index of the flow medium, and Z is the thickness of the schlieren object along the z-direction. According to the Gladstone-Dale equation, the refractive index, n is related to the density, ρ of the flow field of the schlieren object by the equation, $n - 1 = G\rho$, where G is the Gladstone-Dale constant (= 2.23 x $10^{-4}$ $m^3$/kg). In other words, the pixel displacement, $\delta x_{sensor}$, can directly quantify the spatial gradient of the density ($\frac{\partial \rho}{\partial x}$) of the flow field induced by the schlieren object: $\delta x_{sensor} \propto \frac{\partial \rho}{\partial x}$.

## 1.2. Reconstruction methods

There are multiple methods for calculating the pixel displacement and thereby reconstructing the flow fields [29–31]. Consecutive Frame subtraction (CFS) is the simplest BOS visualization method [32]. However, CFS results are only indicative of the flow and do not provide a quantitative relation with density gradients. Here, cross-correlation (CC) and optical flow (OF) techniques are the gold standards. Figure 1d illustrates the basic difference between the CC and the OF techniques. Here, the CC technique is a region-based integral approach whereas the OF is a differential approach. The OF technique can provide a theoretical resolution of up to one vector per pixel. This makes OF particularly interesting for the detection of small displacement values in narrow regions.

Currently, there are several open-source BOS reconstruction software solutions available e.g. PIVLab, OpenPIV, PIVview etc. [20,33,34]. Although originally developed for Particle Image Velocimetry (PIV) applications, these CC techniques are equally suited for BOS reconstructions. Here, the particle displacement between two PIV images (at $t_0$ and $t_0+\Delta t$) is measured by employing CC on small interrogation window (Fig. 1d). To obtain CC coefficients, the interrogation window is swept in the steps of one pixel along both axes in the search area of the flow image. The CC coefficient, *C* can be represented as [35,36]

$$C(x, y) = \sum_{i=1}^{n_x} \sum_{j=1}^{n_y} I_1(i, j) \times I_2(i + x, j + y) \quad \ldots\ldots\ldots\ldots\ldots \text{(1)}$$

where $n_x$, and $n_y$ are the dimensions of the interrogation windows. $I_1$ and $I_2$ are the pixel intensities in the reference and distorted images. The values of *C* are between 0 and 1. Higher values of *C* represent the best match between the image pair, which indicates less distortion in the background. Finally, this yields the displacement values. The interrogation window size affects the resolution and sensitivity of the results. This gives only the sparse displacement that requires proper post-processing. Similarly, the OF technique can also be used as an alternative BOS reconstruction method. It measures the apparent motion of objects between two consecutive frames. Unlike the CC technique, this can provide dense displacements in the whole image. Let's assume, $I(x, y, t)$ and $I(x + dx, y + dy, t + dt)$ are the pixel intensities in the first and second frame captured at time $t$ and $t+dt$, respectively. Also let's further assume that the pixel intensities of an object remain constant between the consecutive frames, and that the same condition can be applied for neighboring pixels as well. Then a fundamental relation of OF can be written as [29,37]

$$\frac{\partial I(x,y,t)}{\partial x}\frac{dx}{dt} + \frac{\partial I(x,y,t)}{\partial y}\frac{dy}{dt} = -\frac{\partial I(x,y,t)}{\partial t} \ldots\ldots\ldots\ldots\ldots\ldots\ldots \text{(2)}$$

where $\frac{\partial I}{\partial x}$, $\frac{\partial I}{\partial y}$, and $\frac{\partial I}{\partial t}$ represent the derivatives of the pixel intensity in the $x$ and $y$-directions and time, $t$ respectively. Here, $\left(\frac{dx}{dt}, \frac{dy}{dt}\right)$ or $(u, v)$ is the image velocity of optical flow and represents the horizontal and vertical components of displacements, respectively. Thus, $(dx, dy)$ gives the displacement field estimated between two consecutive images. The most popular OF algorithms are Lukas-Kanade, Brox algorithm, Horn-Schunck, and Gunnar Farnebäck [38–41]. The methods mentioned above fall into two categories: sparse and dense OF. Sparse OF is based on the motion of distinctive features analysis and is thus not adequate for small displacements. Dense

OF calculates the per pixel displacement and is thus more suitable for BOS reconstruction.

# 2. Methods

## 2.1. Experimental setup

We test Pocket Schlieren with a budget-friendly smartphone (Realme 5 Pro, Make 2019, price: ₹11,999 or $145). This smartphone has a 48 megapixel (MP) Sony IMX586 CMOS imaging sensor, a 6P lens system of 1.79 f/# and 4.73 mm focal-length. In general photography mode, the camera records with a pixel resolution of 1.6 µm where four pixels of 0.8 µm combine to form a large 1.6 µm square pixel. Figure 2a and 2b show the experimental setup to record the density gradient using the smartphone. The distance between camera and schlieren object ($d_1$), as well as the distance between schlieren object and background ($d_2$) must be well optimized for best flow visualization results. For instance, if $d_2$ is large, then even smaller refractive index changes become visible due to the larger pixel displacement, $\delta x$. However, it will mandate reduction in $d_1$ which will further reduce the field of view of the schlieren object seen through the smartphone camera. For our experiment, we used $d_2$ = 0.35 m and the total distance between the smartphone camera and the background ($d_1$+$d_2$) as 0.82 m. The smartphone was fixed rigidly on a stand. We used a white light LED for uniform illumination of the background pattern.

## 2.2. Background pattern:

We have used three different types of background patterns viz. binary dots pattern, binary squares pattern, and grayscale squares pattern in our experiments. All three background patterns were randomly generated and printed on a white A4 size paper. We tested the hot air flow from a burning candle under the same lighting conditions against the three backgrounds and compared the BOS results (see Fig. 3). The signal-to-noise ratio (SNR) of the BOS results for each background pattern has been calculated using the relation $SNR = (I_{sample} - I_{bg})/\sigma_{bg}$ where, $I_{sample}$ is the average intensity of the BOS signal, $I_{bg}$ is the average intensity of the background and $\sigma_{bg}$ is the standard deviation of the background. Our observation from the SNR comparison shows that a random binary squares background pattern performs best.

## 2.3. Pocket Schlieren Android app:

Pocket Schlieren is an Android app that handles the entire BOS process, from image acquisition, to processing, reconstruction, and display of flow visualization on the smartphone. We have used the OpenCV library in our Android app to perform all the image processing tasks. Figure 2c shows the workflow as well as the user interface (UI) of the Pocket Schlieren app. This workflow is split into two branches. The first is for consecutive frame subtraction (CFS) approach and the second for optical flow (OF) approach. Supplementary note 1 further describes the Pocket Schlieren app UI. Even a basic smartphone is powerful enough to provide a live (45-50 FPS at 720p resolution) feed in the CFS mode. The complete processing algorithm flowchart is given in supplementary Fig. 1.

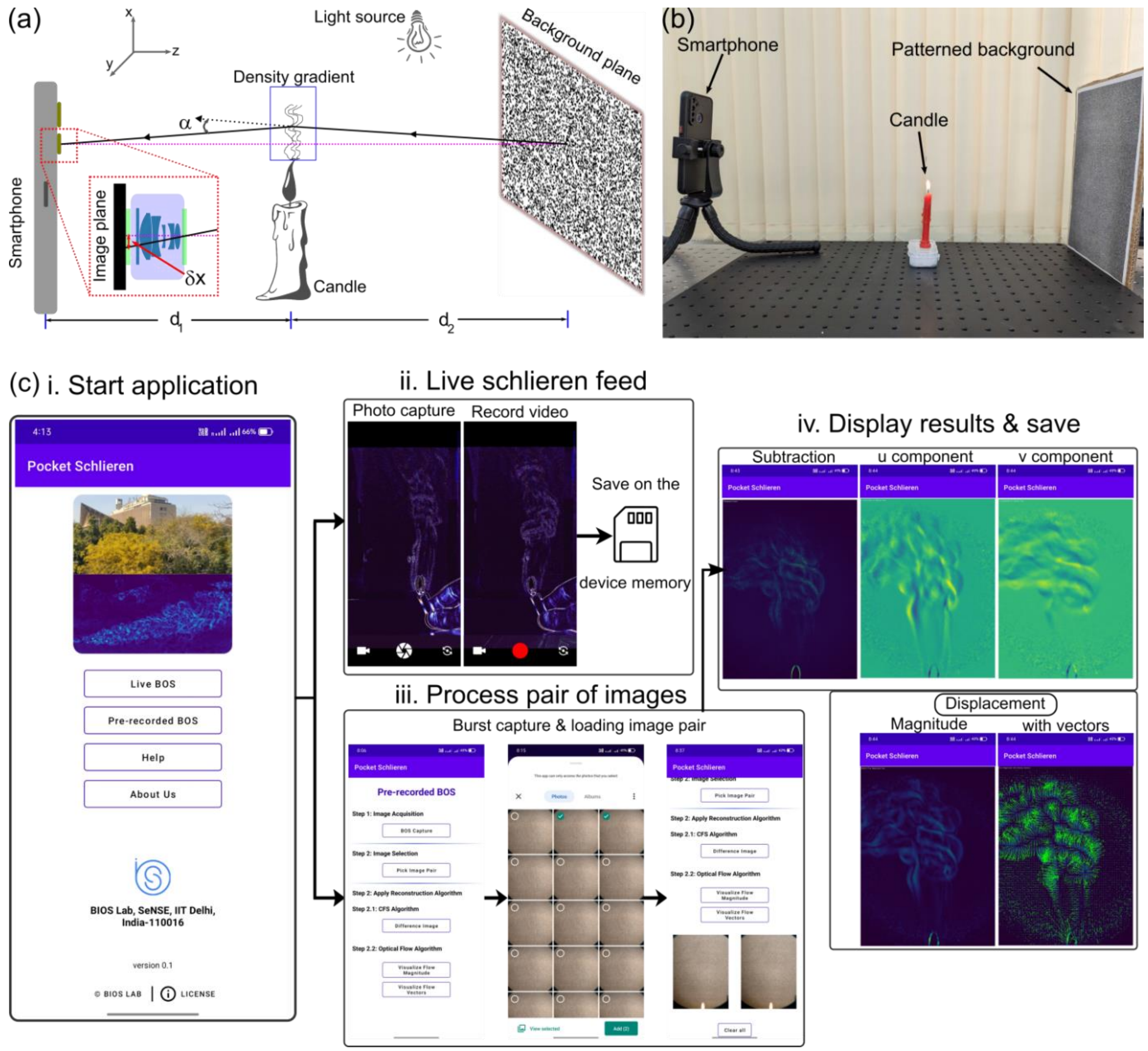

**Fig. 2**: BOS using Pocket Schlieren. (a) schematic for experimental setup, (b) actual experimental setup in lab, (c) UI and workflow of the Pocket Schlieren Android app.

OF is a computationally intensive algorithm. Unlike CFS, OF may not provide rapid processing for a live BOS feed, but it is more powerful to extract smaller density gradients in a flow. We have successfully incorporated the Gunnar Farnebäck OF algorithm in Pocket Schlieren. A practical use of OF in Pocket Schlieren involves 1) a burst mode rapid capture of multiple frames on the smartphone memory, and 2) OF implementation on a pair of captured frames to provide BOS flow visualization. The Gunnar Farnebäck OF is a two-frame motion estimation algorithm and can compute the optical flow for all the points in the frame through polynomial expansion [41]. The workflow of the algorithm is illustrated in the supplementary Fig. 1.

## 2.4. **Data acquisition and analysis**

Pocket Schlieren uses the primary camera system of smartphones for image capture. These are 24-bit (8-bit per channel) RGB color images with JPEG compression. All the image acquisition parameters are set to automatic. Image resolution is limited to 1920 × 1080 pixels to make the on-device computation fast. During our tests the ISO value remained within 100-250, and the exposure time was 10-20 ms. It is possible to use third party apps e.g. Open Camera by Mark Harman [42] for advanced controls on the image acquisition parameters. Pocket Schlieren allows you to load images captured from any third-party hardware or software apps. One can also load images captured by scientific cameras and process them using Pocket Schlieren on a smartphone. All the BOS computations presented here are done on a smartphone using Pocket Schlieren app. Two exceptions from this are shown in the first-row of Fig. 4, where we used PIVlab in Matlab (academic version 2023b), and SNR calculations using ImageJ/Fiji [43,44], both running on a Windows PC.

## 3. Experimental Results

We optimized the background pattern for the Pocket Schlieren experimental setup. Three different types of background patterns were printed on A4 size papers as already discussed in the background pattern section above. After reconstruction of the BOS results, the SNR for each background pattern was calculated for a quantitative comparison. These reconstruction results are shown in Fig. 3.

First, we compare the flow visualization capabilities of CFS and OF approaches in Pocket Schlieren. For this, we recorded plumes from a lighter. We placed the patterned backgrounds: random grey, random dot, and random square one by one at approximately 0.8 m distance from the camera. The lighter was placed at approximately 0.5 m from the camera. To enable systematic comparison, we captured multiple frames in burst mode and used image pairs for both CFS and OF. Figure 3 shows a few frames from this experiment. We observed that a laminar flow, due to smaller density gradients, is hard to visualize in CFS approach. However, OF can detect such small density gradients. On the other hand, CFS approach being very fast is preferred when live visualization is important. Therefore, the default live BOS mode in Pocket Schlieren main UI is based on the CFS approach. We chose OF-based technique to optimize the background patterns for our Pocket Schlieren. We have maintained the same experimental conditions for all the three chosen background patterns. Figure 3 shows the BOS results obtained using our Pocket Schlieren for random gray, random dot and random square pattern, respectively. The quoted SNR values are averaged from eight BOS images. It is evident that the random gray pattern yields the lowest average $SNR_{avg}$ = 12.52 dB. This is due to an overall lower contrast created by gray scale values present in the background pattern here. Random dot background pattern yields moderate average $SNR_{avg}$ = 15.27 dB and the random binary square background pattern shows the highest average SNR value of 25.85 dB. The difference between random dots and random square patterns is due to a better fill fraction control and an efficient camera pixel shape match. Thus, binary patterns offer a distinct upper hand in comparison to grey scale patterns and here binary random square pattern is the best.

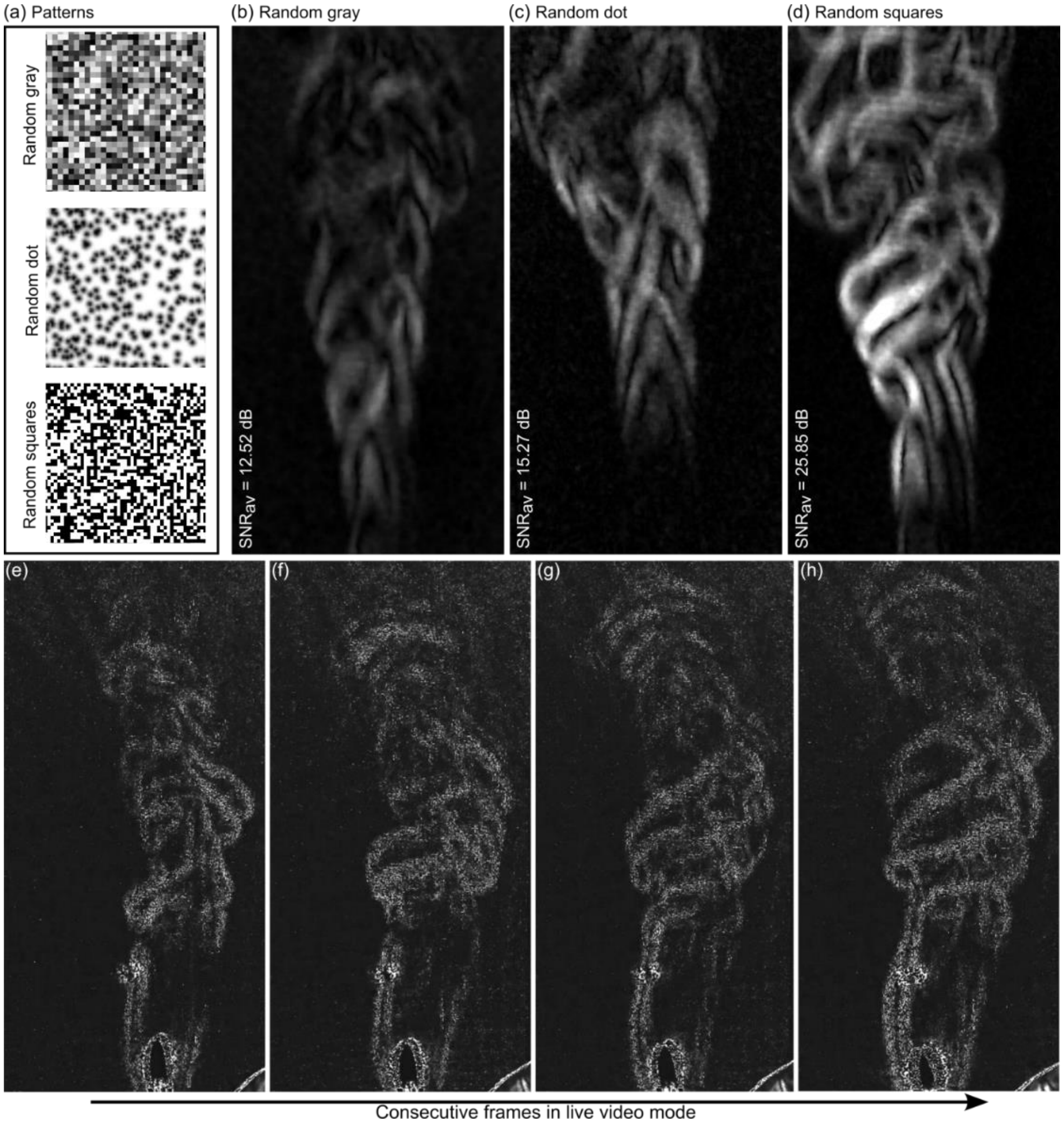

**Fig. 3:** Optimization of the patterned background for Pocket Schlieren. (a) Three patterns under consideration: random gray, random dot, and random squares, respectively, (b)-(d) The reconstructed OF-based BOS results for each of the backgrounds. The noted SNRs are average from eight results, (e)-(h) four consecutive frames recorded using CFS-based live BOS mode. Live BOS performed at 45-50 FPS in 720p resolution. The brightness of the images is enhanced for better visualization.

Now, we perform a comparison of BOS reconstruction between CC based PIVlab and OF based Pocket Schlieren app. PIVlab is a standard PIV software based on MATLAB which is widely used for BOS reconstruction [8]. It is important to see how well OF approach in our Pocket Schlieren performs against PIVlab. Figure 4 shows a side-by-side comparison of BOS reconstruction from both approaches. PIVlab based reconstruction took 5-6 seconds on a Windows desktop (windows 11, Intel Core i5 13th Gen, 16GB RAM). In comparison our Pocket Schlieren reconstructs the BOS results within 10-11 seconds on a smartphone (Realme 5 Pro, Qualcomm Snapdragon 712, 4GB RAM). Figure 4 also shows the comparison of the reconstructed vertical (u component), horizontal (v component), and the net displacement field from the image pair using PIVLab and Pocket Schlieren, respectively. From the zoomed-in region, it is evident that the results obtained from Pocket Schlieren and PIVLab are similar.

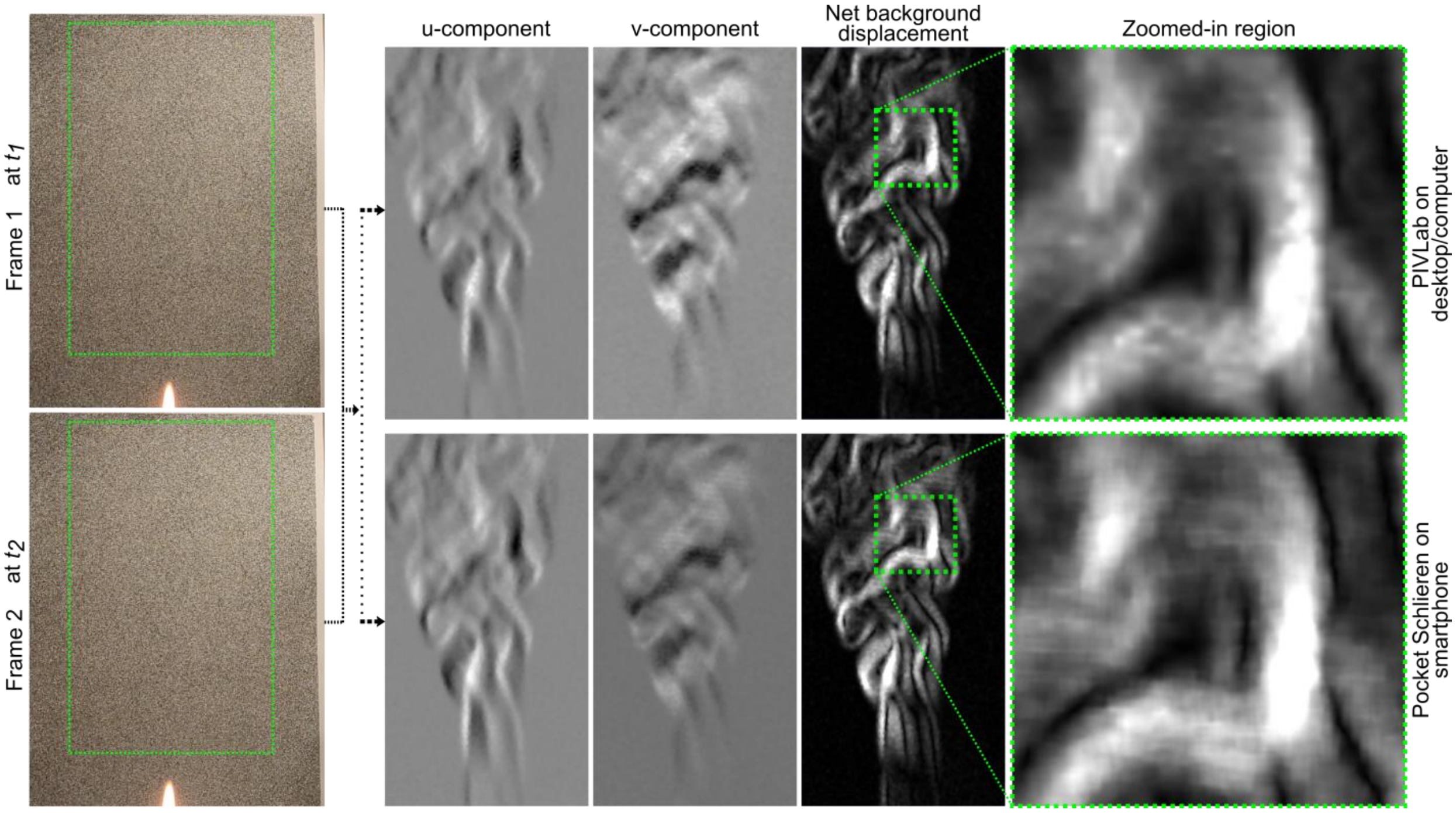

**Fig. 4:** Qualitative comparison of the results obtained from PIVLab and our OF-based Pocket Schlieren app.

Having seen the usability of Pocket Schlieren, next we show results for indoor flow visualization for a variety of schlieren objects. These results are compiled in Fig. 5 where we visualize flow of unburnt butane gas from a lighter, hot air flow from a lit lighter, and convective heat flows in air from a soldering iron, a room heater and an immersion heating rod. The top row in Fig. 5 shows one representative frame captured during each of these experiments. It shows the respective schlieren sources against BOS background patterns.

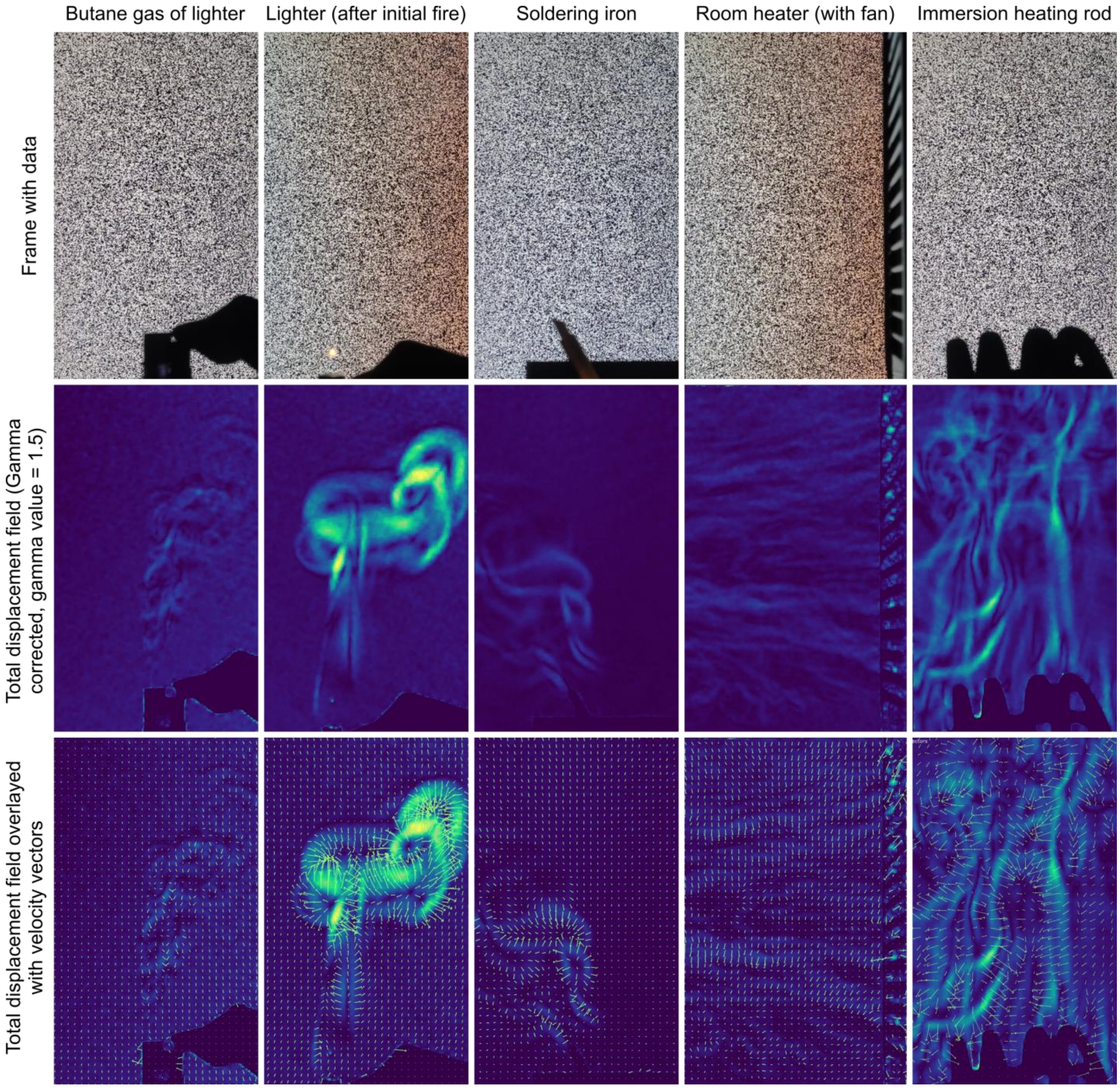

**Fig. 5:** Pocket Schlieren assisted calculation of displacement field and velocity vectors inside flow. Left to right: Different samples used for creating flow which include butane gas lighter, lit lighter, soldering iron, room heater, and immersion heating rod. Top to bottom: a reference data frame, displacement field, and velocity vectors overlayed on top of the displacement field.

The middle row displays respective total background pattern displacement field magnitude obtained from a pair of frames using Pocket Schlieren. The bottom row shows results with background pattern displacement direction overlayed on top of the displacement magnitude. Clearly, Pocket Schlieren can extract weak distortions in the background pattern arising from small density gradients in the air.

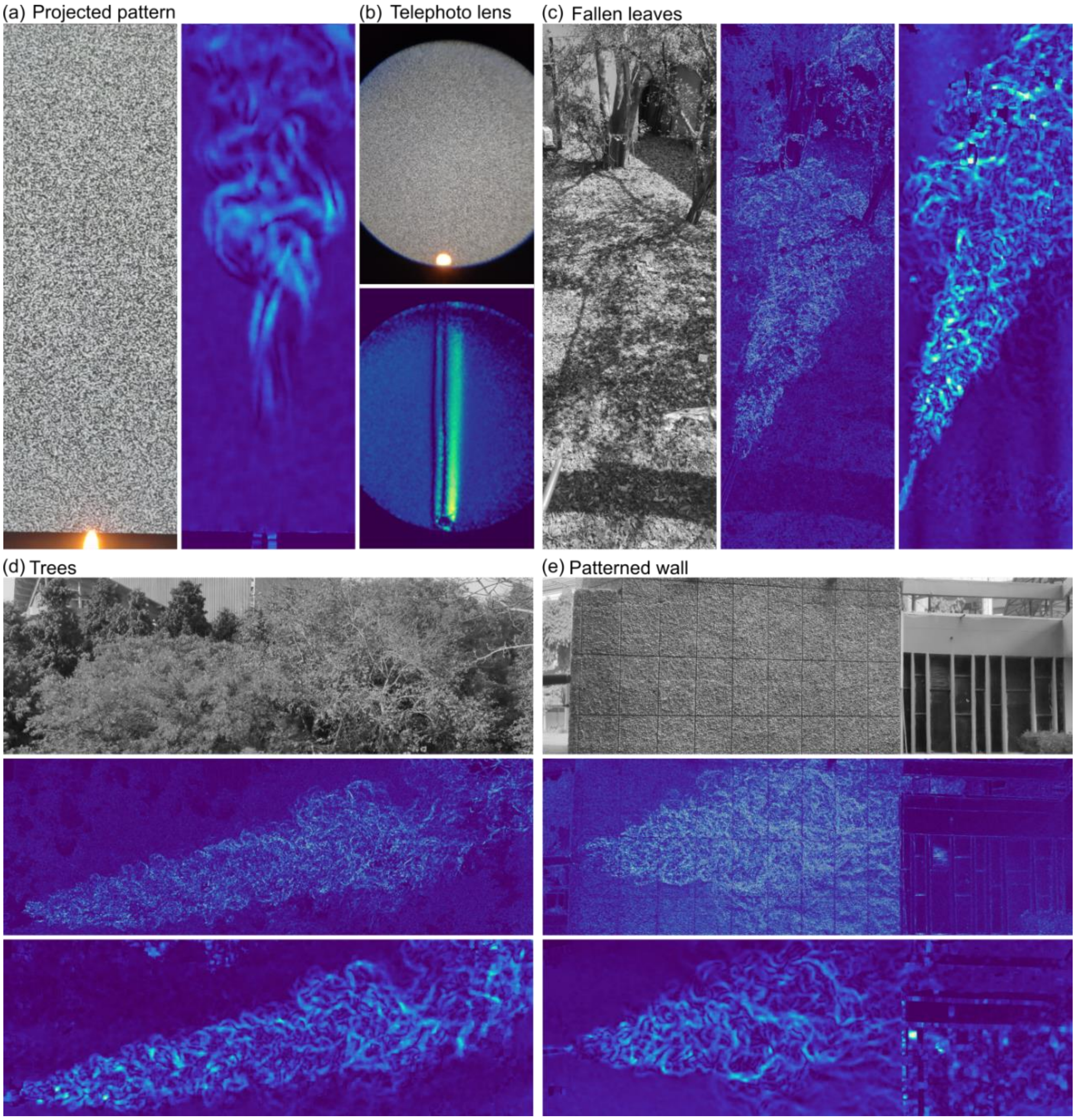

**Fig. 6:** Enhancing sensitivity and field of view with Pocket Schlieren. (a) Flow visualization against a projected background, (b) enhancing BOS sensitivity with an add-on 6x telephoto lens, (c)-(e) large field of view BOS against outdoor natural and manmade backdrops. Fallen leaves in (c), group of trees in (d) and patterned wall in (e). Reconstructed results for both the CFS and OF algorithms are given. CFS and OF results are shown in center and right respectively for (c) & in center and bottom respectively for (d)-(e).

We can easily see the presence of a different gases in air (here butane in air), start of a combustion process (here lighting up the lighter), and convective heat transfer in air (here hot air rising from soldering iron, room heater or immersion heating rod).

Next, we show two sets of results aimed at further enhancing the capabilities of Pocket Schlieren instrument in terms of its sensitivity, and field of view.  The sensitivity of the BOS setup is dependent on the small deflection angle $\alpha$, which is directly related to the distance $d_2$ in Fig. 1a. Thus, a larger separation between the background and the schlieren object ($d_2$) increases the BOS sensitivity. However, this also entails a longer distance between the background and the smartphone camera leading to a very wide field of view imaging. Here, the *Raffel criterion* [7] of 3-5 pixels per background dot will not get satisfied. We solved this problem by adding a 6x telephoto lens in front of the smartphone camera. With this we could image at a significantly larger separation of $d_2$ = 0.7 m and ($d_1$ + $d_2$) = 1.7 m. Figure 6a and 6b show BOS results obtained without and with add-on telephoto lens in our configuration. Here, a projector projected the patterned background onto a white screen (see supplementary Fig. 4a). Figure 6a shows the reconstructed flow field without any add-on lens. The region just above the flame is hottest but corresponding hot air is hardly visible. This is due to a smooth density gradient in the laminar flow. To visualize this small density gradient, we needed to increase the sensitivity of our BOS setup. We do this by adding a 6x telephoto lens in front of the smartphone. With this change, the laminar hot air flow is distinctly visible (see Fig. 6b).

Finally, we demonstrate the ability of Pocket Schlieren to perform large field of view BOS imaging against outdoor and natural backgrounds. Figure 6c-6e show a variety of such BOS imaging results. The original image files of background for each of these results are displayed in the supplementary Fig. 4. These image files are also downloadable from our publicly accessible repository at doi: 10.5281/zenodo.10949271. At first, we utilized ground fallen tree leaves as a patterned background for BOS imaging. We used a burning handheld butane gas torch as the sample. The background was illuminated naturally by the sunlight. The background and torch were approximately 15 m and 0.5 m away from the smartphone. Figure 6c shows an example captured frame in this setup along with flow visualization results using both the CFS and OF algorithm, respectively. OF shows the flow field with high sensitivity and contrast. For the second outdoor demo we used a group of trees as a natural background. Here the group of trees and torch were approximately 25 m and 1 m away from the smartphone. Figure 6d shows results for the flames emanating from the butane torch. As a final outdoor demo example, we chose a patterned gravel wall as the background for BOS capture using Pocket Schlieren. In this demonstration the wall and torch were approximately 6 m and 0.5 m far, respectively. Figure 6e shows the recording and reconstruction of flow from a burning butane torch against this patterned wall background. The flow fields visualization abruptly stops at the right-side boundary of the wall highlighting the importance of the presence of fine structures in the background.

## 4. Discussion

In this paper, we have leveraged the ever-improving hardware features as well as the computational power of modern smartphones and incorporated an entire BOS imaging system called “Pocket Schlieren” on it. It enables on-device reconstruction of the local displacement field, which reduces the complexity, costs, and time. However, there are still a lot of scopes to improve the Pocket Schlieren system. It is evident that information of the background is lost due to the lowpass-filtering behavior of the smartphone camera while projecting it into the image plane. This is due to the averaging of the high-frequency details of the background over one pixel size. This can be, however, mitigated by using wavelet noise pattern as recommended by Atcheson et al. [29].

Although the addition of telephoto lens enhances the BOS sensitivity, it comes at the cost of optical distortion and vignetting in the image. These detrimental optical effects can be removed by using a highly corrected telephoto lens system. Alternatively, in the highly competitive and ever evolving space of smartphones market, newer devices are coming with built-in telephoto functionality. These newer smartphones would be very helpful for high sensitivity Pocket Schlieren applications while retaining the compact form factor and mobility of the tool.

Finally, OF technique is a computationally expensive algorithm, because of which it is hard to implement in real-time. However, it is possible to add the GPU processing power on top of the existing CPU power, to speed up the processing. This will be added in the future release of our Pocket Schlieren.

## 5. Conclusion

Pocket Schlieren is a frugal yet powerful scientific tool for performing BOS imaging. The tool leverages the advanced capturing mode and the computational power of a modern smartphone. In our smartphone BOS imaging system, the local displacement field can be reconstructed from the reference and target image pairs using the OF approach. The home-coded Pocket Schlieren Android app can not only reconstruct the BOS images locally on the device but can also display and capture the BOS events live at video frame rates using a less computationally expensive CFS approach. In our experiments, we have reported the visualization of the local displacement field induced by the heated air due to burning of lighter, blower, soldering iron etc. Pocket Schlieren may be easily extended for sensitive flow visualization, viz. gas leakage and cool flame detection. The simple and interactive UI of the Pocket Schlieren app comes with in-app instructions and resources to become a powerful tool for STEM education and research. This tool does not require any special set-up and is ready to use right out of your pocket even in an outdoor environment. Thus, ‘Pocket Schlieren’ could prove to be a promising scientific tool for mobile visualization of density gradient inside a flow in BOS imaging at an extremely low cost.

## Acknowledgement:

D. Rabha acknowledges the Indian Institute of Technology Delhi for the financial support through Institute Post-Doctoral Fellowship. M. Kumar acknowledges the Indian Institute of Technology Delhi for a seed grant, and the Science and Engineering Research Board, Govt. of India for a startup research grant.

## Author Contributions:

**Diganta Rabha**: Conceptualization, Data curation, Formal analysis, Investigation, Methodology, Software, Validation, Writing. **Vimod Kumar**: Data curation, Validation, Writing. **Akshay Kumar**: Data curation, Validation, Writing. **Dinesh Saini**: Data curation, Validation, Writing. **Manish Kumar**: Supervision, Funding acquisition, Resources, Conceptualization, Data curation, Formal analysis, Investigation, Methodology, Software, Validation, Writing.

# Supplementary Information

## Pocket Schlieren: a background oriented schlieren imaging platform on a smartphone

DIGANTA RABHA[1], VIMOD KUMAR[1], AKSHAY KUMAR[1], DINESH SAINI[2], AND MANISH KUMAR[1*]

**AFFILIATIONS**

*[1]Centre for Sensors, Instrumentation and Cyber-physical System Engineering (SeNSE), Indian Institute of Technology Delhi, Hauz Khas, New Delhi-110016, India*

*[2]Optics and Photonics Centre (OPC), Indian Institute of Technology Delhi, Hauz Khas, New Delhi-110016, India*

*Corresponding author: kmanish@iitd.ac.in

**Supplementary note 1: Pocket Schlieren app details**

Pocket Schlieren app has two BOS modes: 1) CFS and 2) OF. The CFS mode is meant for a fast live BOS visualization. The speed of live BOS comes at the cost of reduced visualization sensitivity of the flow. The CFS UI consists of the following steps:

1) Click on the “Live BOS” button. This opens a new UI page with camera live feed. However, this live feed shows frames with CFS. This same UI page also includes options for recording, image capture and switch camera. These options are available through their respective icons.
2) The image capture icon captures and writes a single CFS image onto smartphone memory.
3) Pressing the record icon writes the CFS live feed onto smartphone memory.
4) The switch camera icon can be used to switch between the front facing and the back facing cameras of the smartphone.

The OF approach is better suited for high sensitivity in the flow visualization. UI of the pocket schlieren app consists of the following steps:

1) Click on the “Pre-recorded BOS” button. It will open a new page containing four buttons for BOS reconstruction process.
2) The “Capture Burst Photo” button captures 10 consecutive photos at once and saves on the smartphone memory. The image resolution is kept at 1920 x 1080 pixels for fast processing and minimal memory requirements. Other parameters, including exposure time, white balance, and ISO, are left in auto mode but remain fixed during a given burst capture sequence.
3) The “Select Pair of Images” button helps in loading two images captured in the previous steps.
4) By pressing the “Visualize Flow Magnitude” button, one can visualize the horizontal and vertical components, as well as the total displacement fields between two frames. Long pressing the results enables a “Save Image” pop-up button to save the results on the device memory. It takes approximately 2 seconds to process 1920 x 1080 pixels resolution images.
5) Clicking the “Flow Visualization Overlaid with Vectors” button provides the velocity vector field overlaid on total displacement field of the flow. It also displays the flow direction of the BOS. It can be saved to the smartphone memory by long pressing the resultant image.
6) The application can also process higher resolution images captured by other devices or apps. It takes approximately 12 seconds to process an image pair of 4000 x 3000 pixels.

The main UI of the pocket schlieren app also has two buttons namely, “Background Pattern Samples” and “Instructions”. Clicking “Background Pattern Samples” button, displays the various background patterns that we used in the experiments of this study. Each pattern has a download option below it so that users may easily store and utilize it right away for experiments. The instructions to use the pocket schlieren app can be accessed by clicking the “Instructions” button.

## Supplementary note 2: Pocket Schlieren app installation and initial set-up

Disclaimer:

Installation and set-up:

1) Download Pocket Schlieren app file from doi: 10.5281/zenodo.10949271 on your android smartphone. The file is named “app_pocket_schliere_v0.1.apk” (file size: 218.6 MB). You may have to ignore a “file may be harmful” message.
2) Touch the APK file in “downloads” folder to install it. A pop-up message will come “Do you want to install app”, you may choose “install”.
3) Install it on your smartphone by touching the APK file.
4) Open the app. The first launch of the app will ask for permissions for camera and microphone access. Allow it.
5) For full functionality, go to “App permission” for the “Pocket Schlieren” app and enable access to camera, files & media, and microphone. Without these permissions the app may crash.
6) Find live BOS image and video captures in internal memory > picture folder > BOSPro folder
7) Find the processing results of “Pre-recorded BOS” in internal memory > picture folder > Burst camera folder.

Next version of the app will improve file handling.

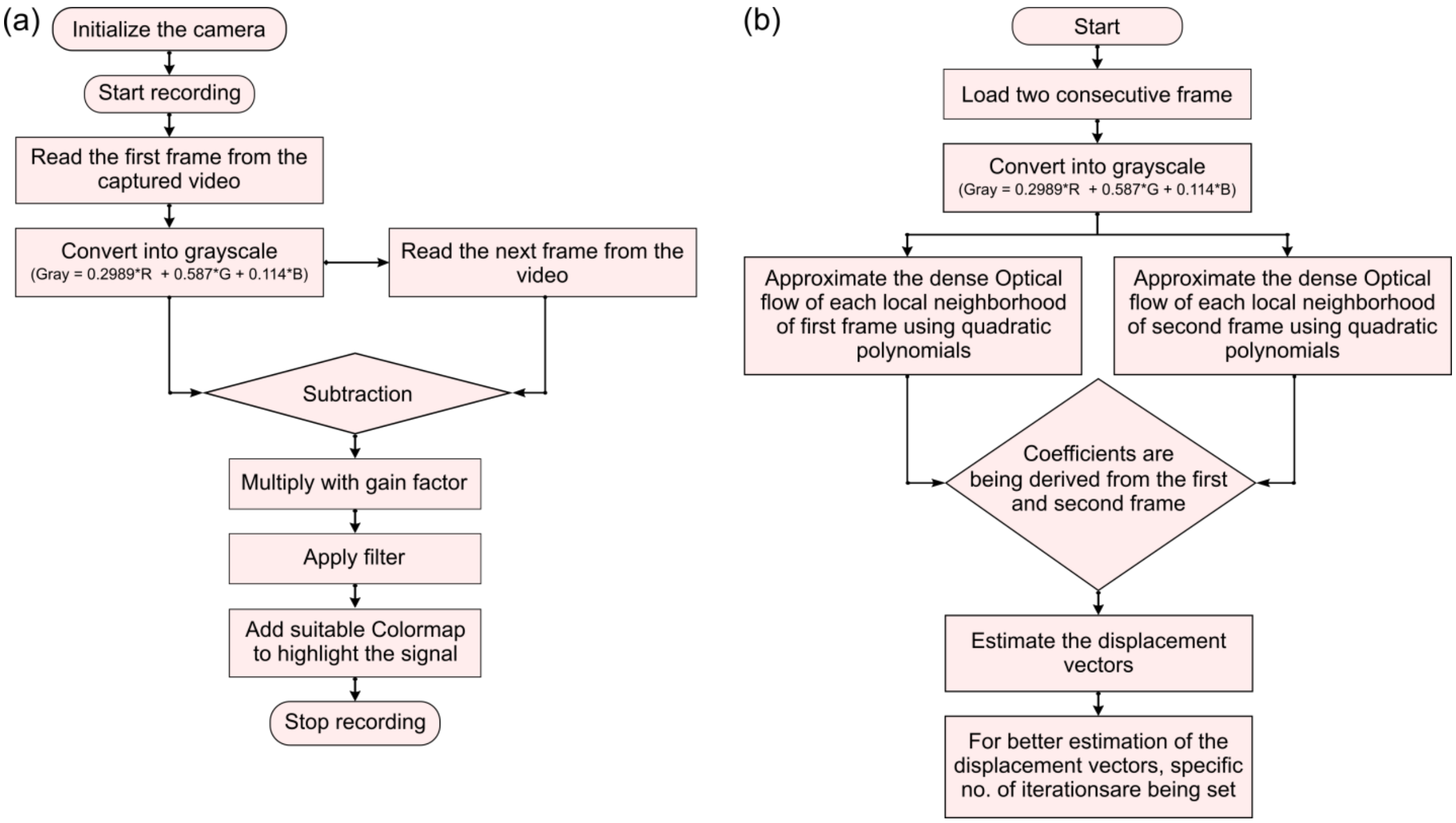

**Supp. Fig. 3:** Pocket schlieren app algorithms. (a) The workflow of the CFS technique, and (b) the Gunnar Farnebäck algorithm.

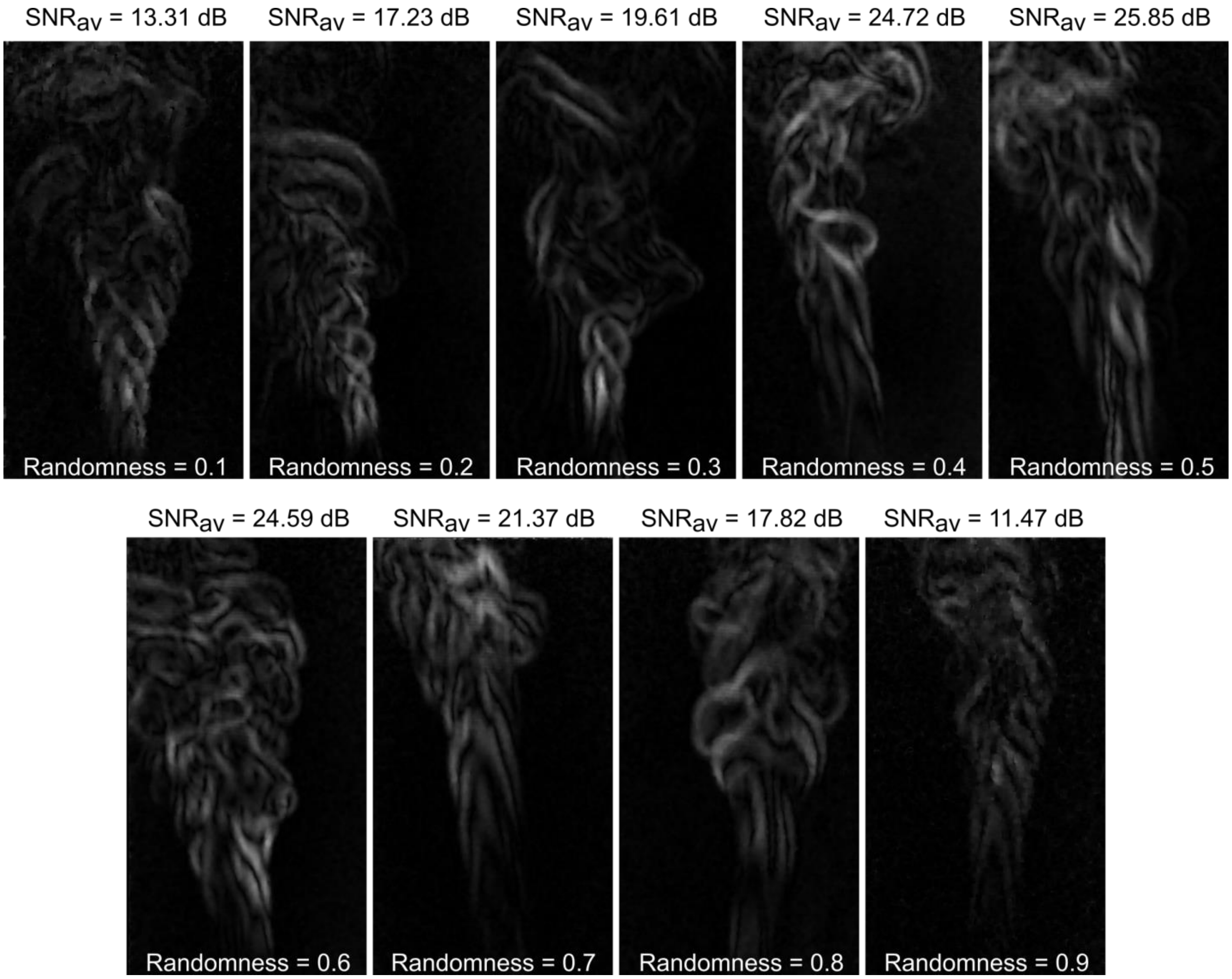

**Supp. Fig. 4:** Optimization of the binary random squares pattern as a background for pocket schlieren.

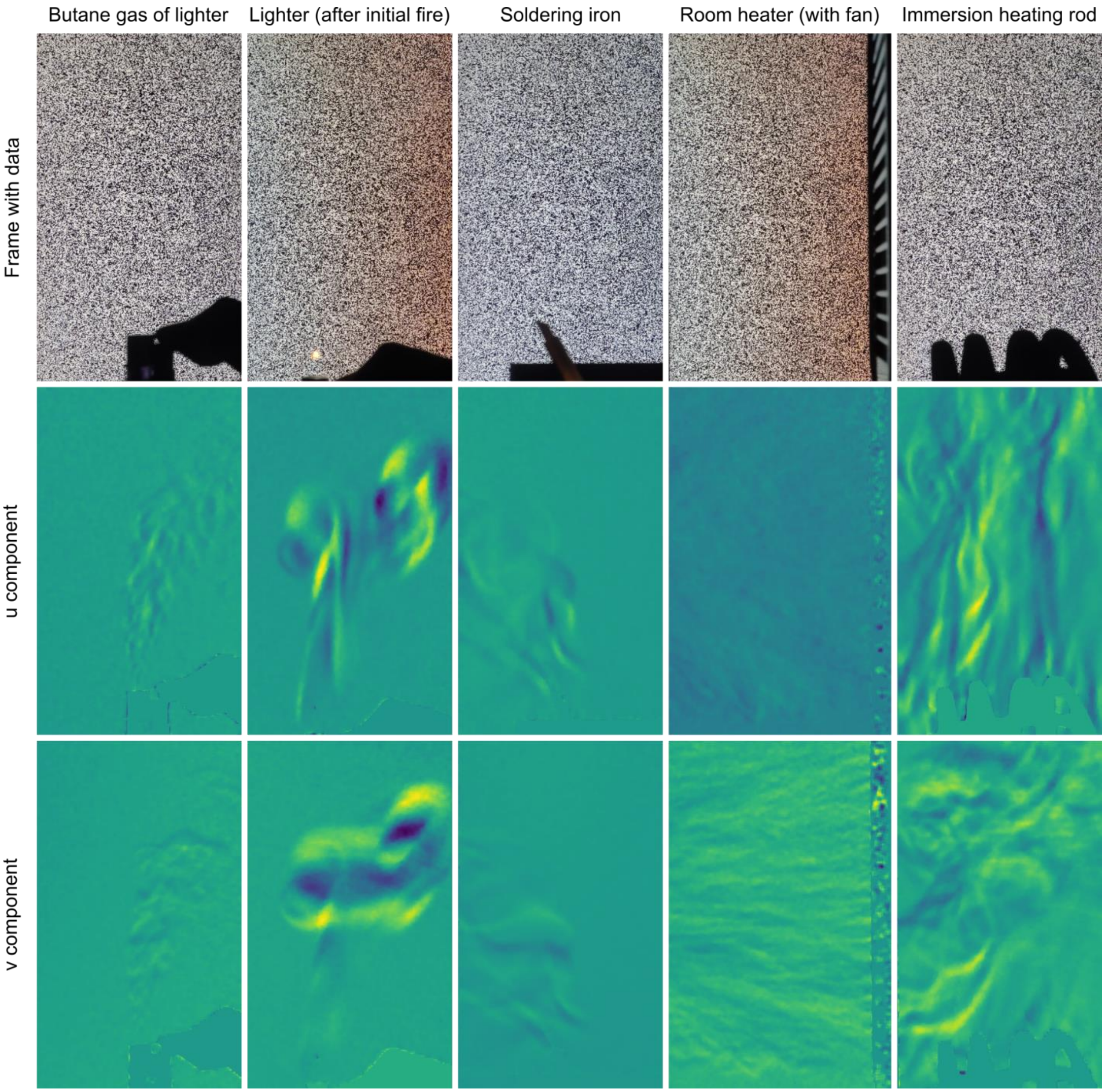

**Supp. Fig. 5:** u- and v-components of the net displacement field reconstructed using the pocket schlieren for the five different schlieren objects- butane gas of lighter, lighter flame, soldering iron, room heater and immersion heating rod, respectively.

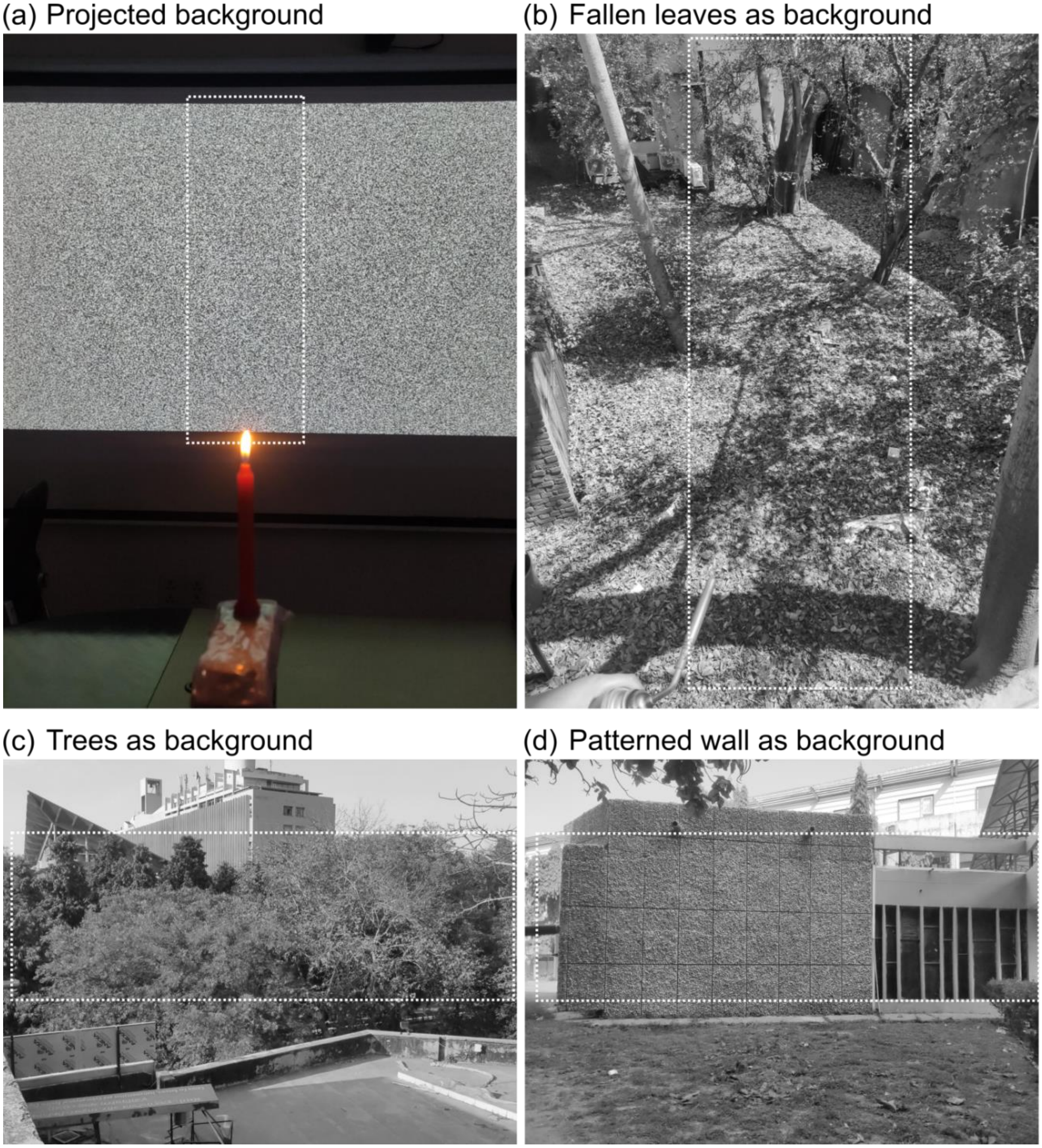

**Supp. Fig. 6:** Ability of the Pocket Schlieren for large field-of-view schlieren imaging against both projected and natural backgrounds. The white rectangle in each image represents the region-of-interest used in the primary manuscript Fig. 6.